\documentstyle[preprint,prc,eqsecnum,aps]{revtex}

\begin{document}
\draft

\title{
Application of Multiple Scattering Theory to Lower Energy \\
Elastic Nucleon-Nucleus Reactions }

\author{ C.R.~Chinn$^{(a),(b)}$, Ch.~Elster$^{(c)}$,
 R.M.~Thaler$^{(a),(d)}$, and S.P.~Weppner$^{(c)}$ \\ ~ \\}
\address{
$^{(a)}$ Department of Physics and Astronomy, Vanderbilt University,
Nashville, TN  37235 \\ ~ \\}

\address{
$^{(b)}$ Center for Computationally Intensive Physics \\
Oak Ridge National Laboratory, Oak Ridge, TN  37831-6373 \\~ \\}

\address{
$^{(c)}$ Institute of Nuclear and Particle Physics,  and
Department of Physics, \\ Ohio University, Athens, OH 45701 \\~ \\}

\address{
$^{(d)}$ Physics Department, Case Western Reserve University,
Cleveland, OH  44106.}

\vspace{10mm}

\date{\today}
\maketitle

\vspace{1in}



\begin{abstract}
The optical model potentials for nucleon-nucleus elastic scattering
at $65$~MeV are calculated for $^{12}$C, $^{16}$O, $^{28}$Si,
$^{40}$Ca, $^{56}$Fe, $^{90}$Zr and $^{208}$Pb in first order
multiple scattering theory, following the prescription of the
spectator expansion, where the only inputs are the free NN
potentials, the nuclear densities and the nuclear mean field as
derived from microscopic nuclear structure calculations.
These potentials are used to predict
differential cross sections, analyzing powers and spin rotation
functions for neutron and proton scattering at 65 MeV projectile
energy and compared with available experimental data.
The theoretical curves are in
surprisingly good agreement with the data.  The modification of
the propagator due to the coupling of the struck nucleon to the
residual nucleus is seen to be significant at this energy and
invariably improves the congruence of theoretical prediction and
measurement.
\end{abstract}

\pagebreak


\narrowtext


\section{Introduction}
\hspace*{10mm}
Recently measurements of 65~MeV neutron-nucleus elastic differential
cross sections have been published \cite{65mev}. Together with the
corresponding proton data, one now has an opportunity to study the
effects of isospin degrees of freedom and to analyze multiple scattering
theory with and without the coulomb field at low energies.
In the past it had been assumed that first order multiple
scattering theory would prove unable to provide an accurate
representation of experiments at
such low energies and thus models in the form
of phenomenological optical potentials or effective
nucleon-nucleon interactions were proposed.  With recent advances
in our ability to calculate the first term in a multiple scattering
expansion together with the the influence of the mean field potential
binding the struck nucleon to the target nucleus,
it is now possible to
test whether one is able to represent low-energy scattering in
the first order or whether one must consider higher order scattering
terms, all within a parameter-free description.

\section{Formalism}

\hspace*{10mm}
We have been pursuing a program of calculation of elastic
nucleon-nucleus scattering at energies sufficiently high such that
first order multiple scattering theory in the forward cone
provides a good description of the data \cite{medium,chinn}.
In this program the only inputs are the nucleon-nucleon (NN)
interaction, represented by NN t-matrices, the target wave
functions, and the static target nuclear mean field.  These quantities
are incorporated into multiple scattering theory in the hierarchical
spectator expansion for the optical potential, in which the transition
operator, $T$, is defined to be
\begin{equation}
T = U + U G_0 P T . \label{eq:2.1}
\end{equation}
The operator $P$ is the projector onto the target initial state.
The optical potential operator $U$ is given in the spectator
expansion as:
\begin{equation}
U = \sum_i \tau_{i} + \sum_{i,j\neq i} \tau_{ij} + \ldots ~ .
\label{eq:2.2}
\end{equation}
The first order theory corresponds to a truncation of this
series to a single term,
\begin{equation}
U \approx \sum_i \tau_{i} ~ , \label{eq:2.3}
\end{equation}
where
\begin{equation}
\tau_{i} = v_{0i} + v_{0i} G_0 \left(1-P\right) \tau_{i} \label{eq:2.4}
\end{equation}
Here $v_{0i}$ stands for the potential between the projectile nucleon ($0$)
and the $i$th target nucleon.  The propagator $G_0$ is given by
\begin{eqnarray}
G_0^{-1} & = & E - h_0 - H_A \nonumber \\
 & = & E - h_0 - h_i - \sum_{j\neq i} v_{ij} - H^i \nonumber \\
 & = & E - h_0 -  H_i - H^i , \label{eq:2.5}
\end{eqnarray}
where $h_i$ is the kinetic energy operator for nucleon $i$ and
$H_A = H_i+ H^i$ is the target Hamiltonian.
The operator, $\tau_i$, can be reexpressed in a solvable
one-body integral equation as
\begin{equation}
\tau_i = {\widetilde t_{0i}} - {\widetilde t_{0i}} G_0 P \tau_i ~,
\label{eq:2.6}
\end{equation}
where ${\widetilde t_{0i}}$ is given to be:
\begin{eqnarray}
{\widetilde t_{0i}} & = & t^{free}_{0i} + t^{free}_{0i}
  \left[ G_0 - g_0 \right] {\widetilde t_{0i}}  \nonumber \\
  ~& = & t^{free}_{0i} + t^{free}_{0i} g_0 {\cal T}_i g_0
{\widetilde t_{0i}}  ~. \label{eq:2.7}
\end{eqnarray}
Here $t^{free}_{0i}$ is the free NN t-matrix and $g_0$ is the free NN
propagator for the active pair consisting of projectile and
target nucleon $\left[g_0^{-1} = E' - h_0-h_i\right]$.
The scattering operator ${\cal T}_i$ expresses the scattering of
the target nucleon ($i$) from the residual nucleus, which is
represented by $\sum_{j\neq i} v_{ij}$.  The explicit treatment of
Eq.~(\ref{eq:2.7}) is described in detail in Ref.~\cite{medium}
and is directly derivable within the spectator expansion of
multiple scattering theory.  The right term of
Eq.~(\ref{eq:2.7}) results from the difference between the free
propagator $g_0$ with $G_0$, which corresponds to the propagation
of the target nucleon through the nuclear medium and can be
thought of as a propagator modification.

The first order optical potential is then constructed with the
operator, $\tau_i$, from Eq.~(\ref{eq:2.6}):
\begin{equation}
U_{opt} = \langle{\vec k'}_0\Psi_A| \sum_i \tau_i
  |{\vec k}_0\Psi_A\rangle ~. \label{eq:2.8}
\end{equation}
In the present calculations, which are performed in momentum space,
$U_{opt}$ enters in the `optimum factorized' or
`off-shell $\tau\rho$' form as
\begin{equation}
U_{opt} \approx \tau(q,{\cal K};E) \rho(q) ~, \label{eq:2.9}
\end{equation}
where $q=k_0'-k_0$ and ${\cal K}=\frac{1}{2}\left(k'_0+k_0\right)$;
$k_0'$ and $k_0$ are the final and initial momenta
of the projectile.
This corresponds to a  steepest descent evaluation of the
`full-folding' integral, in which the $\tau$ is convoluted with the
nonlocal density as indicated schematically in
Eq.~(\ref{eq:2.8}). For harmonic oscillator model densities it has
been shown that the optimum factorized form represents the nonlocal
character of $U_{opt}$ qualitatively in the intermediate energy
regime \cite{FF,FFC}. Complete `full-folding' calculations with
more realistic nuclear densities are in progress.
It is to be understood that all spin summations are performed in
obtaining $U_{opt}$ (under the usual assumption of a spin-saturated
target), thus reducing the required NN t-matrix elements to  the
spin-independent component (corresponding to the Wolfenstein
amplitude $A$) and the spin-orbit component (corresponding to
the Wolfenstein amplitude $C$).
All scattering calculations presented here contain an additional factor
in the optical potential to account for the transformation of the NN
t-matrix from the two-nucleon c.m. frame to the nucleon-nucleus c.m.
frame \cite{pttw}.

\section{Theoretical Predictions}

\hspace*{10mm}
For a calculation of the first order optical potential in the
optimum factorized form, the quantities
$\tau(q,{\cal K};E)$ and $\rho(q)$ are required as input.
All calculations presented in this paper are based on the full Bonn
Potential \cite{Bonn} as the NN interaction from which $\tau(q,{\cal
K};E)$ is obtained.
As can be seen from Eq.~(\ref{eq:2.6}), the quantity
($\tau\rho$) can be calculated as the solution of a one-body
integral equation in which (${\widetilde t}\rho$) serves as
the driving term.  The proton densities
are taken from charge densities measured in electron scattering
experiments \cite{electron}.  Although
the neutron distribution in nuclei are not completely determined
by measurement, nuclear structure calculations indicate
significant differences between the proton and neutron distributions.
These differences are nonnegligible in our reaction calculations
and especially affect the location of the diffraction minima
and become even more pronounced in the case of neutron
scattering.  We believe neutron densities taken
from the Hartree-Fock-Bogolyubov calculations of Ref.~\cite{HFB}
give the best presently available representation of the neutron
distributions.  These are the densities used in the present work.

The calculation of $\widetilde t_{0i}$ according to Eq.~(\ref{eq:2.7})
requires the free NN t-matrix as well as ${\cal T}_i$, the `t-matrix'
representing the scattering of the struck target nucleon from the
residual nucleus. A one-body mean field potential for a nucleon within
the target nucleus is used as the driving term to obtain ${\cal T}_i$ as
solution of a Lippmann-Schwinger type equation. The inclusion of this
correction, ${\cal T}_i$, corresponds to a modification of the free NN
propagator, $g_0$, due to the nuclear medium in order to recover
the propagator $G_0$, which correctly represents the propagation of the
target nucleon through the nucleus.
Our calculations  use two different models for the  mean fields, one is
the nonlocal, nonrelativistic mean field potential taken from a
Hartree-Fock-Bogolyubov calculation \cite{HFB}. Curves based on this
choice are represented by dash-dotted lines.
The second choice involves a nonrelativistic
reduction of the  mean field potentials resulting from a Dirac-Hartree
calculation based upon the $\sigma$-$\omega$ model \cite{DH}.
Curves based on this choice are represented as dashed lines.
Calculations using only the free NN t-matrix correspond to a truncation
of Eq.~(\ref{eq:2.7}) after the first term, so that
\begin{equation}
{\widetilde t}_{0i} \approx t^{free}_{0i} ~. \label{eq:2.10}
\end{equation}
Results using this truncation are included in the figures to illustrate
the importance of the correction due to  ${\cal T}_i$ in the energy
regime under discussion. The corresponding curves are shown as solid
lines.

\section{Comparison with Data}

\hspace*{10mm}
The elastic neutron and proton elastic scattering observables
[differential cross section $\frac{d\sigma}{d\Omega}$, analyzing
power $A_y$ and spin rotation function $Q$]
are calculated for $^{12}$C,
$^{16}$O, $^{28}$Si, $^{40}$Ca, $^{56}$Fe, $^{90}$Zr and $^{208}$Pb.
In all of these cases, except for $^{16}$O and $^{90}$Zr, neutron
differential cross section data \cite{65mev} exist for the natural elements
along with the more extensive proton data.  The carbon
calculations in Fig.~1 should not be taken as seriously as the others, since
it is known that carbon is a highly deformed nucleus.  For a
realistic calculation of carbon certain additional
collective degrees of freedom need to be considered.

The overall impression is that the first order, parameter free
multiple scattering
predictions provide a very good representation of the
data for angles below $\sim 60^\circ$, which is a very pleasant
surprise.
One possible partial explanation for this good agreement is that the
momentum transfer for $65$~MeV for $\theta=180^\circ$ is roughly
equivalent to that corresponding to $\theta=60^\circ$ at $260$~MeV.
That is to say, the range of momentum transfer under observation is
rather small relative to that considered at higher energies.
Nonetheless, this good a description of
the data is unexpected.  At this low an energy, the higher order
terms in the spectator expansion were expected to become
important.  It now appears that these terms will not be strongly
structured and so will, no doubt, raise the forward cross section
and fill in the sharp diffraction minima. A change in the spin
structure due to higher order terms may only be apparent at large
scattering angles.

In a recent publication the authors concentrated on the total
cross section $\sigma_{tot}$ for neutron scattering from
$^{16}$O and $^{40}$Ca as
a function of scattering energy \cite{neutron1}.  It was found
that our calculations reproduced the  neutron total cross section
data above $\sim 100$~MeV, whereas at
 65~MeV the first order multiple scattering theory predictions
for the total cross sections for $^{40}$Ca fall short of the
measured values by about
$\sim 10\%$.  Thus, for example, if one looked at the proton
elastic differential cross section in the forward cone, one might
be surprised to see how well the data is represented by the
theoretical predictions.  On the other hand, it is observed that
the theoretical curve for the neutron elastic differential cross
section consistently falls below the data in the forward cone.
The reason is that in the proton case the Coulomb effect dominates
at small scattering angles and thus masks any underprediction.
It may also be argued that the neutron projectile should be able to
penetrate further into the nucleus than a proton projectile, due
to the lack of the coulomb barrier.  This may perhaps mean that for
heavier nuclei, neutron scattering might be more sensitive to higher
order multiple scattering effects as well as to the medium
corrections used here. This is offered to illustrate the enhanced value
and impact of having combined neutron and proton data sets.

The curves presented show very clearly that the
inclusion of the propagator modification arising from the coupling
of the struck nucleon to the residual nucleus improves the
description of the data, in some cases dramatically.  For neutron
scattering only differential cross section data for angles up to
60 degrees are at present available.
While this data is limited, a definite improvement from
the propagator modification is seen.  Relative to the unmodified
calculation the diffraction minima systematically move to higher
angles in better accord with the observation and the diffraction
minima become less sharp also in accord with observation.
For proton scattering, extensive data are available for
$\frac {d\sigma}{d\Omega}$, $A_y$ and $Q$.  While there is always
marked improvement in the differential cross section description
due to the propagator medium modification, the effect on
the spin observables $A_y$ and
$Q$ is even more striking.  For $A_y$ at the first diffraction
minima the propagator modification moves the theoretical curve closer
towards the data so that for up to about $50-60$~degrees there is a
pretty good representation of the data.  The effect on the $Q$
function is even more dramatic where the medium correction
shifts the curve down significantly so that the theoretical curve
sits almost directly on top of the data for angles up to about
$60$~degrees.  It appears that for $Q$, the propagator
modification is required to obtain an excellent agreement with the
data,  and it is further remarkable that this is the case for all nuclei
under consideration.  The theoretical predictions for the heavier
elements $^{90}$Zr and $^{208}$Pb are not as good as for the lighter nuclei,
but
the dramatic effect observed for the $Q$ function remains.
The effect of the propagator modification
on the spin rotation is slightly larger for neutron scattering
than for protons and in Figs.~4-7 this difference is even larger
for both $Q$ and $A_y$.
The good description of the spin observables
independent of the mass of the nucleus may be taken as evidence that
at these low energies the scattering process is generally completely
surface dominated, and by construction our theoretical approach captures
the effect of the coupling of the struck target nucleon to the mean
field of the residual nucleus.

All the nuclei under consideration are  even-even
nuclei.  However, in a shell model sense they are mostly not
`spin-saturated', and so one might expect that all Wolfenstein
parameters in the NN t-matrix to contribute
to the optical potential.  The remarkable description of the data
which is obtained here based on  only $A$ and $C$ suggests that such
effects are small.  This may give information
concerning the structure of the target,  but in the present stage it
would be premature to test sensitivities to the correlation structure of
the target.

\section{Summary and Conclusion}

The success of our calculations at higher energies \cite{chinn,neutron1}
have emboldened us to venture toward energies lower than
previously considered.  Neutron and proton elastic scattering
observables are calculated in first order multiple scattering theory
in a parameter free fashion for a number of even-even spin zero
targets at $65$~MeV.  Even at this low
energy a good description of the data is obtained.  It is observed
that the propagator modification due to the coupling of the struck
nucleon to the residual nucleus invariably brings the theoretical
results into closer agreement with observation.

The availability of both neutron and proton scattering data at
the same energy is especially valuable. At this energy we find
that our theoretical predictions
consistently underpredict the neutron differential cross section in the
forward direction; whereas, in the corresponding proton differential cross
section, Coulomb effects mask this underprediction. This illustrates the
enhanced value and impact of having available combined neutron and proton
data sets.


\vfill
\acknowledgments
The authors would like to thank R.W.~Finlay and J.~Rapaport
for many discussions concerning this work. The computational
support of the the Ohio Supercomputer Center under
Grants No.~PHS206 and PDS150 is gratefully acknowledged.
This work was performed in part under the auspices of the U.~S.
Department
of Energy under contracts No. DE-FG02-93ER40756 with Ohio University,
DE-AC05-84OR21400 with Martin Marietta Energy Systems, Inc., and
DE-FG05-87ER40376 with Vanderbilt University.  This research has also
been supported
in part by the U.S. Department of Energy, Office of Scientific Computing
under the High Performance Computing and Communications Program (HPCC)
as a Grand Challenge titled ``The Quantum Structure of Matter".


\pagebreak


\pagebreak
\noindent
\begin{figure}
\caption{The differential cross section, analyzing power and spin
        rotation functions are shown for elastic nucleon scattering
        from $^{12}$C.  The left three panels show elastic neutron
        scattering, while the right three panels are for proton
scattering.  The
        neutron data are from Ref.~\protect\cite{65mev}, while the proton
        data are from Ref.~\protect\cite{c12}.
        In all cases, the solid line corresponds to a calculation
 of the first order optical potential based on the free NN t-matrix from
the full Bonn model \protect\cite{Bonn}
as input. The dashed and dash-dotted lines include the propagator
modification due to the nuclear mean field. For the dash-dotted curve a
Hartree-Fock-Bogolyubov \protect\cite{HFB} is used, for the dashed curve a
Dirac-Hartree mean field \protect\cite{DH}.  \label{fig1} }
\end{figure}

\begin{figure}
\caption{The same as Fig.~1, except for $^{16}$O and the
        elastic proton scattering data are from Ref.~\protect\cite{ca40}. }
\end{figure}

\begin{figure}
\caption{The same as Fig.~1 except for $^{26}$Si and the
        elastic proton scattering data are from Ref.~\protect\cite{ca40}. }
\end{figure}

\begin{figure}
\caption{The same as Fig.~1 except for $^{40}$Ca and the
        elastic proton scattering data are from Ref.~\protect\cite{ca40}. }
\end{figure}

\begin{figure}
\caption{The same as Fig.~1 except for $^{56}$Fe and the
        elastic proton scattering data are from Ref.~\protect\cite{ca40}. }
\end{figure}

\begin{figure}
\caption{The same as Fig.~1 except for $^{90}$Zr and the
        elastic proton scattering data are from Ref.~\protect\cite{ca40}. }
\end{figure}

\begin{figure}
\caption{The same as Fig.~1 except for $^{208}$Pb and the
        elastic proton scattering data are from Ref.~\protect\cite{ca40}. }
\end{figure}

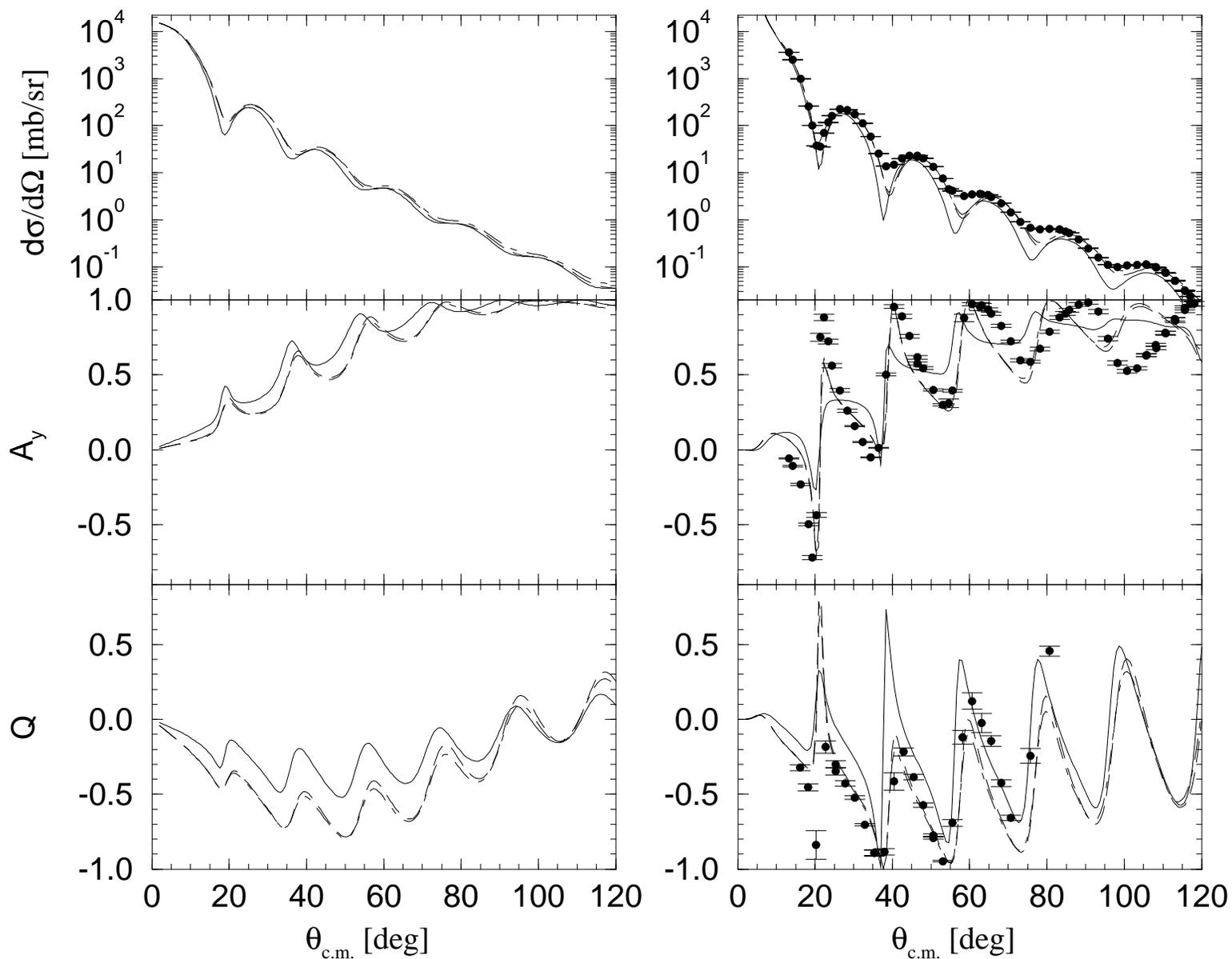
\end{document}